\documentstyle[aps,prl,preprint]{revtex}
\draft

\begin{document}


\title{Bounds on the decay of the auto-correlation in
phase ordering dynamics.}

\author{ Chuck Yeung$^{1,2}$,
Madan Rao$^{3}$ and Rashmi C.\ Desai$^{1}$}

\address{
$^{1}$Department of Physics, University of Toronto,
Toronto, Ontario M5S-1A7 CANADA\\
$^{2}$Division of Science,
Pennsylvania State University at
Erie,  The Behrend College,
Erie, PA, 16563 USA\\
$^{3}$Institute of Mathematical Sciences,
C.I.T.\ Campus, Taramani, Madras 600113 INDIA
}

\date{\today}

\maketitle

\widetext                               

\begin{abstract}

We obtain bounds on the decay exponent $\lambda$ of the
autocorrelation function in phase ordering dynamics (defined by
$\lim_{t_{2} \gg t_{1}}
\langle \, \phi( {\bf r}, t_{1} ) \phi( {\bf r}, t_{2} ) \rangle \sim
L(t_{2})^{-\lambda}$).  For non-conserved order parameter,
we recover the Fisher and Huse inequality, $\lambda \geq d/2$.
If the order parameter is conserved we also find $\lambda \geq
d/2$ if $t_{1}$ = 0. However, for
$t_{1}$ in the scaling regime, we obtain $\lambda \geq d/2 + 2$ for
$d \geq 2$ and $\lambda \geq 3/2$ for $d=1$.  For the one-dimensional scalar
case, this, in conjunction
with previous results, implies that $\lambda$ is different for $t_{1} = 0$
and $t_{1} \gg 1$. In 2-dimensions, our extensive numerical
simulations for a conserved scalar order parameter show that
$\lambda \approx 3$ for $t_{1}=0$ and  $\lambda \approx 4$ for $t_{1}
\gg 1$.  These results contradict
a recent conjecture that conservation of order parameter requires
$\lambda = d$.
Quenches to and from the critical point are also discussed.

\end{abstract}

\bigskip



Phase separation dynamics proceeds when a system is quenched from its high
temperature, homogeneous phase to a low temperature, inhomogeneous phase
(where several phases coexist in equilibrium).
Due to its simple
description yet rich behavior, phase ordering dynamics has greatly
enhanced our understanding of non-equilibrium processes \cite{REVIEWS}.
At late times,
the spatial distribution of domains can be described by a single
time-dependent length, $L(t)$ which typically grows algebraically
in time, $L(t) \sim t^{1/z}$.  This results in a scale invariant
equal-time correlation function $C(r,t)$. More recently it has been realized
that the unequal-time correlation function is scale covariant. In particular,
the asymptotic decay of the two-time autocorrelation
function, $C(r,t_{1},t_{2}) = \langle \, \phi( {\bf r}, t_{1} ) \phi( {\bf 0},
t_{2}) \rangle$ defines an independent exponent $\lambda$, via
$\lim_{t_{1} \ll t_{2} }
C(0,t_{1},t_{2}) \sim ( L(t_{1})/L(t_{2}) )^{\lambda}$. This exponent
bears no apparent
relation to the growth exponent $z$ and so its value provides a sensitive
test for approximate theories of phase ordering kinetics
\cite{FISHER,FURUKAWA,TWOTIME,NEWMAN,BRAYLIC,LIUAUTO,HUMAYUN}.
Although the autocorrelation function
has been studied extensively for non-conserved order parameter
dynamics\cite{FISHER,TWOTIME,NEWMAN,BRAYLIC,LIUAUTO,HUMAYUN,BRAY1,MASON},
there has been hardly any work on conserved dynamics.
However
in a recent Letter, Majumdar et al.\ have shown numerically and
analytically that
that $\lambda = 1$ for $m=1$, $d=1$ and $t_{1} = 0$ where
$m$ is the number of components in the order parameter \cite{MAJUMDAR}.
It has been further argued\cite{MAJUMDAR,MAD} that {\it the conservation of
order parameter demands that $\lambda = d$ for all $m$}.

In this Letter, we obtain lower bounds on the decay exponent $\lambda$.
For non-conserved order
parameters, $\lambda \geq d/2$ independent of $t_{1}$, consistent with a
general argument of Fisher and Huse \cite{FISHER}.  For conserved
order parameters, we also obtain $\lambda \geq d/2$ for $t_{1} =
0$ (assuming the quench is from a high temperature phase).  However,
for $t_{1}$ in the scaling regime, we find that $\lambda \geq d/2 +
2$ for $d \geq 2$ and $\lambda \geq 3/2$ for $d=1$.
This difference arises from
the small $k$ behavior of the scattering
intensity $S(k,t_{1})$.
In conjunction
with the exact result  for $\lambda$, for the 1-dimensional
scalar model, with $t_{1}=0$ \cite{MAJUMDAR}, we conclude that
for $d=1$, $\lambda$ depends on whether $t_{1}=0$ or $t_{1} \gg 1$.
To carry out the investigation in higher dimensions, we
perform an extensive numerical integration of the
Cahn-Hilliard equation (see Eq.\ (\ref{eq:CH}) below) in $d=2$.  We find that
$\lambda \approx 3$ for $t_{1} = 0$ and $\lambda \approx 4$ for $t_{1}$
in the scaling regime.  {\it This is
inconsistent with the recent conjecture that
$\lambda =d$}  \cite{MAJUMDAR,MAD}.
We discuss why this conjecture fails. We
also derive bounds on $\lambda$ for quenches to and from the critical point.
Our results easily extend to vector order parameters.

We begin by  obtaining the lower bounds on $\lambda$.
The equal point auto-correlation $ C( t_{1}, t_{2} ) \equiv
C({\bf 0}, t_{1}, t_{2} )$ is related to the
$k$ space auto-correlation $S(k,t_{1},t_{2})$ by
$$
        C(t_{1},t_{2}) =
                \int d {\bf k} \;
        \langle \, \delta \phi_{{\bf k}}(t_{1})
        \,\delta \phi_{-{\bf k}}(t_{2}) \, \rangle
        =
        \int d {\bf k}  \; S( k, t_{1}, t_{2} ).
$$
Here $\phi( {\bf r}, t)$ is the order parameter at point ${\bf r}$
and time $t$ and $\delta \phi( {\bf r}, t ) \equiv \phi( {\bf r}, t )
- m_{0}$ with $m_{0} = V^{-1} \int d {\bf r} \; \phi( {\bf r}, t )$
and the fourier transform $\delta \phi_{\bf k}(t) \equiv V^{-1/2} \int d
{\bf r} \; e^{-i {\bf k} \cdot {\bf r} } \delta \phi( {\bf r}, t )$.
The angular brackets indicate an average over initial conditions.
For a critical
quench $\langle \, m_{0} \rangle = 0$ and $\langle m_{0}^{2} \rangle$
is ${\cal O}( V^{-1} )$, whereas for an off-critical quench $\langle \, m_{0}
\,
\rangle$ is ${\cal O}(1)$.

Using the Cauchy-Schwartz inequality, we find
\begin{eqnarray}
        C(t_{1},t_{2})
        & \leq &
         \int \, d {\bf k} \;
        \langle \, \delta \phi_{{\bf k}}(t_{1})
         \,\delta \phi_{-{\bf k}}(t_{1}) \, \rangle ^{1/2}\,
          \langle \, \delta \phi_{{\bf k}}(t_{2})
        \,\delta \phi_{-{\bf k}}(t_{2}) \, \rangle ^{1/2},
                \nonumber \\
        & \sim &
         \int \, d {\bf k} \;
         S(k,t_{1})^{1/2} \, S(k,t_{2})^{1/2}.
         \label{eq:BOUND}
\end{eqnarray}
\noindent
where  $S(k,t) = S(k,t,t)$.

Now assume $t_{2}$ to be in the scaling regime with $t_{2} > t_{1}$.
At late times, the scattering is due to the sharp interfaces or
defects. The $k$ modes, $\delta \phi_{{\bf k}}$ at times $t_{1}$ and $t_{2}$
will be uncorrelated when the interfaces move a distance greater than
$2 \pi/k$ so that
$S(k, t_{1}, t_{2} )$ decreases rapidly for
$k \, ( \, L(t_{2}) - L(t_{1}) \, ) \gg 1$.
The upper limit of the integral over
$k$ in Eq.\ \ref{eq:BOUND} can then be cut off at $2 a \pi/L(t_{2})$
where $a$ is a constant of ${\cal O}(1)$\cite{EXPLICIT}.
For $L( t_{2} ) \gg L(t_{1})$,
only the small $k$ behavior of $S(k,t_{1})$ contributes to the
integral.
Assume that $\lim_{k \to 0 } S(k,t_{1}) \sim k ^{\beta}$ ($\beta \geq 0$).
For quenches to zero temperature, $S(k,t_{2})$ will have the
scaling form $S(k,t_{2}) = L(t_{2})^{d} f( k L(t_{2}) )$.
Substituting  into Eq.\ (\ref{eq:BOUND})
(with the appropriate limits of integration), gives
\begin{eqnarray}
        \lim_{t_{2} \gg t_{1} } C(t_{1}, t_{2} )
        \sim    L(t_{2})^{-\lambda}                     
        & \leq &
        L(t_{2})^{d/2}
        \int^{ 2 a \pi/L(t_{2})}_{0} d k \; k^{d-1} \;
        k^{\beta/2} f( k L(t_{2}) ),
                \nonumber \\
        & \sim &
        L(t_{2})^{-(d+\beta)/2}\,.
        \nonumber
\end{eqnarray}
This immediately gives a lower bound on $\lambda$,
$$
        \lambda \geq \frac{ \beta + d }{2}.
$$
The argument just presented, holds for conserved and
nonconserved, scalar and vector order parameters.

We now consider specific dynamical scenarios. Let $T_{I}$ and $T_{F}$
be the
temperatures of the initial and final states respectively. We first
focus on quenches from the high temperature phase ($T_{I}=\infty$) to zero
temperature ($T_{F}=0$). Since the initial state is disordered,
$\lim_{k \to 0}
S(k,0) \sim k^{0}$. In the absence of a conservation law,
$\lim_{k \to 0}
S(k,t_{1}) \sim k^{0}$ for both $t_{1} = 0$ and $t_{1}$ in the scaling
regime. Therefore $\beta = 0$ and
\begin{equation}
        \lambda \geq d/2.
                \label{eq:BOUND1}
\end{equation}
This inequality was also obtained by Fisher and Huse
using general scaling arguments \cite{FISHER}
and is consistent with all
results to date\cite{TWOTIME,NEWMAN,LIUAUTO,HUMAYUN,BRAY1,MASON}.
For $t_{1} = 0$, conservation
of the order parameter does not affect this inequality
since $\beta=0$ for $t_{1}=0$. However, if $t_{1}$ is
in the scaling regime, then $\lim_{k \to 0}
S(k,t) \sim k^{4} $ for $d \geq 2$ \cite{YEUNG}
and $\beta = 4$.
For $d = 1$, the dynamics is dominated by noise and
Majumdar et al.\ find that $\lim_{k \to 0} S(k,t) \sim k^{2}$ \cite{MAJUMDAR},
so that $\beta = 2$ for $d=1$. Therefore, for $t_{1}$ in the scaling regime,
\begin{equation}
        \lambda \geq  \begin{array}{ccc}
        \frac{d}{2} + 2 & \; & \mbox{if $d \geq 2$},\\
        \frac{3}{2} & & \mbox{if $d=1$}.
        \end{array}
                \label{eq:BOUND2}
\end{equation}
These bounds suggest that the asymptotic exponent {\it may} depend on whether
$t_{1}$ is, or is not in the scaling regime
but do not rule out
that the exponent is independent of $t_{1}$.
However for $d=1$, Majumdar et al.\ find analytically
and numerically that $\lambda = 1$ for $t_{1} = 0$, while we find that
$\lambda \geq 3/2$ for $t_{1}$ in the scaling regime\cite{CAUTION}.

For vector fields (with $m$, the number of components of the order parameter,
$ > 2$), an argument analogous to Ref.\,\cite{YEUNG}, gives the same
$\lim_{k \to 0} S(k,t) \sim k^{4}$.
This is supported by an extensive numerical integration of the Cahn-Hilliard
equation\cite{MRAC}. Therefore the lower bounds on $\lambda$
derived above are valid even for vector order parameters with $m > 2$.

Quenches from the critical point ($T_{I}=T_{c}, T_{F}=0$) lead to long-range
correlations of the initial configurations. In this case,
$\lambda \geq d/2$ no longer  holds.  More generally
if $S(k,0) \sim k^{-\sigma}$ we obtain $\lambda \geq (d-\sigma)/2$. (For
critical dynamics $\sigma = 2 - \eta$ where $\eta$ is the static critical
exponent).  This is consistent with the result of Bray et al.\ who
found that, for nonconserved order parameter,
 $\lambda = (d-\sigma)/2$ for $\sigma$ greater than a critical
value $\sigma_{c}$ \cite{BRAYLIC}.

Analysis of the bounds on the autocorrelation exponent for quenches {\it to}
the critical point ($T_{I}=\infty, T_{F}=T_{c}$), has to start afresh from
Eq.\ (\ref{eq:BOUND}). Since $t_{2}$ is in the critical point scaling regime,
the correlation function has the following scaling form,
$S(k,t_{2}) \sim k^{-2+\eta} f_{c}(kL(t_{2}))$.  Substituting this form
into Eq.\ (\ref{eq:BOUND}) gives $\lambda \geq (2d-2 +\eta + \beta)/2$.
Therefore when $t_{1} = 0$ we get $\lambda \geq (2d-2 +\eta)/2$.
When $t_{1}$ is also in the scaling regime, the
bound on $\lambda$ depends on the behaviour of the scaling function
$f_{c}(kL(t_{1}))$ as $kL(t_{1}) \to 0$. For
nonconserved systems $\lim_{x \to 0} f_{c}(x) \to const.$\,, or
$\beta = -2 + \eta$ leading to $\lambda \geq d-2+\eta$.

These lower bounds on $\lambda$ of course do not fix the value of the
exponent. As previously mentioned, exact analytical and numerical
computations on the 1-dimensional scalar model have been carried out for the
case when $t_{1}=0$.
In higher dimensions, however, the empirical results are not very conclusive
\cite{FURUKAWAN}.  We therefore compute the asymptotic value of $\lambda$ by
numerically integrating the
Cahn-Hilliard equation in two-dimensions,
\begin{equation}
        \frac{ \partial \phi( {\bf r}, t ) }{ \partial t }
                =
        \nabla^{2} \mu( {\bf r}, t ),
                \label{eq:CH}
\end{equation}
where $\mu = -\phi + \phi^{3} - \nabla^{2} \phi$.
We used an Euler discretization with
$\delta t = 0.1$ and $\delta x = 1.09$ and periodic boundary
conditions. We
discretize the Laplacian as
$$
        \nabla^{2} \phi_{i,j} = \frac{1}{ \delta x^{2} }
                \frac{ \sqrt{2} }{ 1 + \sqrt{2} }
                \left[ \frac{1}{2} \sum_{n.n.n.}
                + \sum_{n.n.} - 6 \right] \phi_{i,j}.
$$
This choice decreases lattice anisotropy effects and allows a
larger $\delta t$ before the onset of the checkerboard instability
\cite{OONO}.
This dynamical equation is solved
subject to random initial conditions which are uncorrelated and uniformly
distributed between $-0.05$ and
$0.05$ (the initial state is disordered).
Decreasing $\delta t$ has no effect on the numerical results.  Increasing
$\delta x$ to $1.32$ results in pinning effects which lead
to a slower
decay of the autocorrelation function at late times (even though
the effect on the single-time behavior is less apparent).

We have used the interfacial area density as a measure of the characteristic
lengthscale $L(t)$.  Operationally, this is defined as $( 2 \, \delta
x \, n_{x} n_{y} )/ n_{opp}$, where $n_{x} n_{y}$ is the total number of
lattice sites and
$n_{opp}$ is the number of sites with a nearest neighbor with $\phi$
of opposite sign.  We recover the standard result that $L(t)$ grows
as $t^{1/3}$ for all $t > 400$.  Other measures of the characteristic
lengthscale, such as the first zero of the real space correlation function,
also behave in the same manner (for $t > 400$).

\begin{figure}
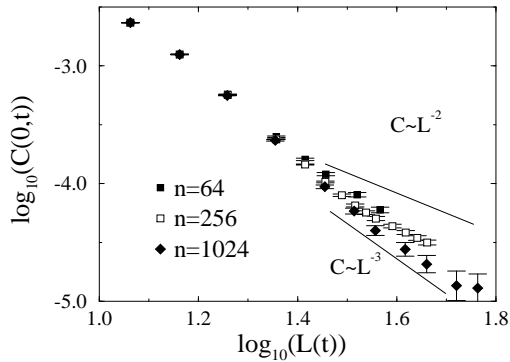


\caption{
\label{fig:AUTO}
The autocorrelation function $C(0,t)$ for $n=64$ (3084 initial conditions),
$256$ (1120 initial conditions) and $1024$
(42 initial configurations).  Lines corresponding
to $C(0,t) \sim L^{-3}$ and $C(t) \sim L^{-2}$ are shown for
comparison.
}

\end{figure}

Fig.\ \ref{fig:AUTO} shows $C(t_{1},t)$ vs.\ $L(t)$ for $t_{1} =
0$ and $t$ between 100 and 12800 for three lattice sizes
$n=n_{x}=n_{y} = 64$ (3084 initial conditions), $n=256$ (1120 initial
conditions)
$n=1024$ (42 initial conditions).  Concentrating on the
largest lattice size, $C(0,t) \sim  L^{-3.7}$ for approximately
two decades of $t$ in its decay.  There is a crossover to a slower
decay at late times with $C(0,t) \sim L^{-3.0}$. At extremely late times
there is an indication of an even slower decay, which we
attribute to finite size effects.

\begin{figure}
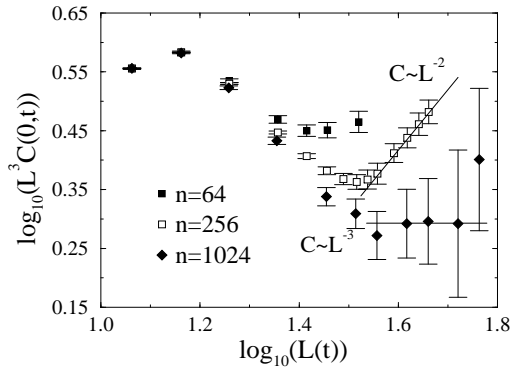


\caption{
\label{fig:AUTOL3}
$L^{3} C(0,t)$  vs.\ $L(t)$ for same data as Fig.\ 1.  Lines corresponding
to $C(0,t) \sim L^{-3}$ and $C(t) \sim L^{-2}$ are shown for comparison.
These results emphasize that the slower decay at late times is due to finite
size effects.
}

\end{figure}

To emphasize the asymptotic trend,
Fig.\ \ref{fig:AUTOL3} shows $L^{3} C(0,t)$ vs.\ $L(t)$ for the
same data.  Here it is clearer that
the slower
late time decay occurs at earlier times for smaller lattices, indicating
finite size effects.
The importance of the finite size effects
was initially surprising since, for single time quantities, finite size
effects only become important when $L(t)$ is of order of the lattice
dimension, $L_{0}$.
Thus the usual length scales extracted from
single-time quantities  were identical for $n=256$ and $n=1024$.
However, since $C(0,t)$
decays rapidly with $t$,  any small systematic effect
becomes
increasingly relevant as $t$ increases.
Clearly finite size effects on $C(0,t)$
can be important (though not necessarily so) when
the spread in $C(0,t)$ is of the same order as $C(0,t)$.
The spread in $C(0,t)$ decreases as $L_{0}^{-d/2}$ and, based on our
simulations, depends only weakly on $L(t)$.
Hence finite size effects can become important when
$C(0,t) \sim L(t)^{-\lambda} \sim L_{0}^{-d/2}$, i.e.,
much earlier than for single-time quantities.
>From  Fig.\ \ref{fig:AUTOL3}, $C(0,t)$ for $n=64$ first shows
significant differences from the $n=256$ result at $C \approx
2.5 \times 10^{-4}$ or $L \approx 18$.  Thus we expect
the $n=1024$ data to be free of finite size effects down to
$C \approx 2.5 \times 10^{-4}/16 \approx 1.5 \times 10^{-5}$ or
$L \approx 50$.  Therefore we make the preliminary conclusion
that the true asymptotic value of
$\lambda$ is approximately $3$.  Finally note that the late time
result for $n=256$ is consistent with $\lambda = d = 2$.  However,
comparing this with the result for $n=1024$ indicates that this regime
is due to the finite size of the lattice.

\begin{figure}
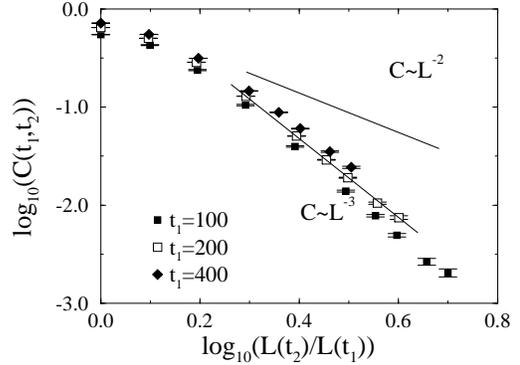


\caption{
$C(t_{1},t_{2})$  vs.\ $L(t_{2})/L(t_{1})$ for $n=1024$ and $t_{1} =
100$, $200$ and $400$. Lines corresponding to $C(t_{1},t_{2}) \sim L^{-4}$
and $C(t_{1},t_{2}) \sim L^{-2}$ are shown for comparison.
\label{fig:T1}
}

\end{figure}

Fig.\ \ref{fig:T1} shows $C(t_{1}, t_{2})$
vs.\ $L(t_{2})/L(t_{1})$ for $t_{1}  =$ 100, 200 and 400.  Although we
cannot rule out a further slower
decay, our result is consistent with $\lambda \approx 4$ for
$t_{1}$ in the scaling regime.
Hence,
we find that, for the $d=2$, conserved, scalar model,
$\lambda \approx 3$ for $t_{1} = 0$ and
$\lambda \approx 4$, for $t_{1}$ in the scaling regime.
To pin down these values more precisely would require simulations
on larger lattices and for larger values of $t_{1}$ and $t_{2}$.

Our numerical results for $d=2$ and the lower bound on $\lambda$
for $t_{1}$ in the scaling regime (Eq.\ (\ref{eq:BOUND2})) are inconsistent
with the recent conjecture
(Refs.\cite{MAJUMDAR,MAD}) that  conservation of order parameter requires that
$\lambda=d$ in all cases.  We believe the apparent inconsistency is
because
the argument leading to $\lambda = d$, incorrectly
applies a scaling analysis
to the non-scaling, ${\bf k} = 0$ mode.  To be explicit,
we briefly review the argument of Ref.\cite{MAJUMDAR} (the argument presented
in Ref.\cite{MAD} is similar).  The Cahn-Hilliard equation (Eq.\
(\ref{eq:CH})) in $k$ space is
$$
        \frac{ \partial \phi_{{\bf k}}(t) }{ \partial t }
        =
        D( {\bf k}, t),
$$
where $D( {\bf k}, t ) = k^{2} \mu_{k}(t)$.  Define
$\tilde{S}(k,0,t) \equiv \langle \, \phi_{{\bf k}}(0)
\phi_{-{\bf k}}(t) \, \rangle = S(k,0,t) + V \delta_{k,0}
\langle m_{0}^{2} \rangle$.  The formal solution to $\tilde{S}(k,0,t)$
is
$$
        \tilde{S}(k,0,t)
        =
        \tilde{S}(k,0,0)
        \exp\left( \int^{t}_{0} dt' \;
        \frac{ \gamma(L(t'), k L(t') ) }{ t' }
        \right),
$$
where $\gamma(L(t), k L(t) ) \equiv
t \; \langle \, D({\bf k},t) \phi_{-{\bf k}}( t) \, \rangle/
\tilde{S}(k,0,t)$.
In the scaling regime, $\gamma(L, kL) = \gamma(kL)$ and we obtain
(in the limit $t \gg 0$)\,,
\begin{eqnarray}
        \int^{t}_{0} dt' \;
        \frac{\gamma(k L(t') )}{t'}
        & = &
        z  \int^{kL(t)}_{kL(0)} dx \;
        \frac{ \gamma( x ) }{ x },
              \nonumber  \\
        & = &
        z  \gamma(0) \log\left( \frac{ L(t) }{ L(0) } \right)
        + z  \gamma_{1} \left( k L(t) \right),
                \nonumber
\end{eqnarray}
where $L(t) \sim t^{1/z}$ and $\gamma_{0} = \lim_{x \to 0^{+}} \gamma(x)$.
The result is
$$
        \tilde{S}(k,0,t) =
        \tilde{S}(k,0,0)
        \left( \frac{L(t)}{L(0)} \right)^{\gamma_{0} z}
        F_{1}( k L(t) ).
$$
Since $\int d {\bf k} \; \tilde{S}(k,0,t) = \langle \, \phi({\bf
r}, 0) \phi( {\bf r}, t) \, \rangle \sim L(t)^{-\lambda} =
L(t)^{-d+\gamma_{0} z}$ or $\lambda = d - \gamma_{0} z$.  However,
since $\tilde{S}(k=0,0,t)$ is constant in time (conservation law !),
it is argued that
$\gamma_{0}$ must vanish and hence $\lambda = d$.

Our contention is that,
even though
the relation
$\lambda = d - \gamma_{0} z$  with $\gamma_{0} = \lim_{x \to 0^{+}}
\gamma(x)$ holds,  the conclusion that, due to conservation of order
parameter, $\gamma_{0}$ necessarily vanishes does not.  This is
because the relation $\lambda = d - \gamma_{0} z$ is based on scaling,
which only holds for
$k > 0$, while the vanishing of $\gamma_{0}$ is based
on the conservation of
order parameter which only
holds for ${\bf k} = 0$.
Therefore the conclusion $\gamma_{0} = 0$ is invalid since it
applies a scaling argument to the nonscaling ${\bf k} = 0$ behavior.

To see this more clearly, consider the quasi-static scattering intensity
$\tilde{S}(k,t)=\tilde{S}(k,t,t)$.  For large $t$, $\tilde{S}$ is given
by $\tilde{S}(k,t) = L(t)^{d} f( k L(t) ) + V \delta_{k,0} \langle \,
m_{0}^{2} \, \rangle$.  Global conservation of order parameter requires
that  $\tilde{S}(k=0,t)
=$ constant, while the locally conservative dynamics requires
that $f(0) = 0$.  Hence global conservation
leads to a discontinuity at $k=0$ for both
$\tilde{S}(k,t_{1},t_{2})$ and $\gamma(L,kL)$ \cite{NOTE}.
The singularities in
$\gamma(x)$ and $\tilde{S}(k,t_{1},t_{2})$
at  $k=0$ can be removed by choosing the initial
distribution so that $\langle \, m_{0}^{2} \, \rangle = 0$.
However,
in this case, $\tilde{S}(0,t_{1},t_{2})$ vanishes {\it independent of the value
of $\gamma_{0}$}, so that the application of the conservation
law does not fix the value of $\gamma_{0}$. (A.\ J.\ Bray has made a similar
argument proving $\lambda=d$ does not necessarily hold\cite{BRAYPC}).

Having provided useful lower bounds on $\lambda$,
we now ask whether it is possible to bound $\lambda$ from above?
Unfortunately, we have not been able to provide useful upper bounds.
However, we note that the bound (Eq.\ (\ref{eq:BOUND2})) as well as
our numerical results violate the upper bound conjectured by Fisher and
Huse,
$\lambda \leq d$ \cite{FISHER}.  As they originally noted,
this conjecture contains many assumptions. Moreover, in as much as their
argument is aimed at the decay of the magnetization, their conjecture has
validity only when the order parameter is {\it not} conserved and when $t_{1} =
0$ (so that $S(k,t_{1}) \sim k^{0}$).

We thank David Jasnow for the use of his workstations.  We are grateful
to John Ross and the University of Toronto Instructional and Research Computing
Centre
(UTIRC). This work was partially
supported by the NSERC Canada.

\begin{references}
\bibitem{EXPLICIT}
The explicit
assumption is that $S(k,t_{1},t_{2})$ decays faster than
$(k L(t_{2}))^{-d}$.
\bibitem{CAUTION}
$S(k,t) \sim k^{4}$ holds exactly only  as $t \to \infty$.
The second bound $\lambda \geq (d+4)/2$  holds for
$\lim_{t_{2}/t_{1} \to \infty} \lim_{t_{1} \to \infty}$.
For large but finite
$t_{1}$ there is a large range of $t_{2}/t_{1}$ for which
the bound holds with the range increasing with $t_{1}$.
\bibitem{NOTE}
For the non-conserved order parameter
there is no singularity at $k=0$.
\end{references}
\end{document}